\documentclass[12pt,journal,compsoc]{IEEEtran}

\usepackage{amsmath}
\usepackage{amsfonts}
\usepackage{amssymb}

\usepackage{float}
\usepackage{tikz}
\usepackage{times}
\usepackage{url}
\usepackage{multirow}
\usepackage{color}
\usepackage{textcomp}

\usepackage{graphicx}
\usepackage{graphics}
\usepackage{graphicx}
\usepackage{epstopdf}
\usepackage{ed}
\usepackage{wrapfig}
\usepackage{subfig}

\usepackage{booktabs}
\usepackage{algorithmicx}
\usepackage{algorithm}
\usepackage{algpseudocode}
\floatstyle{ruled}
\newfloat{algorithm}{tbp}{loa}
\floatname{algorithm}{Algorithm}
\usepackage{graphics}
\usepackage{graphicx}
\usepackage{epstopdf}
\usepackage{ed}
\usepackage{wrapfig}
\usepackage{subfig}

\usepackage{booktabs}

\usepackage[T1]{fontenc}
\usepackage{float}
\usepackage{amstext}
\usepackage{verbatim}
\usepackage{textcomp}

\usepackage{algorithm}
\usepackage{algpseudocode}
\floatstyle{ruled}
\newfloat{algorithm}{tbp}{loa}
\floatname{algorithm}{Algorithm}

\algnotext{EndFor}
\algnotext{EndIf}

\usepackage{graphicx}

\usepackage{mdwlist}
\usepackage{paralist}

\usepackage[section]{placeins}

\usepackage{algorithmicx}
\usepackage{algorithm}
\usepackage{algpseudocode}
\floatstyle{ruled}
\newfloat{algorithm}{tbp}{loa}
\floatname{algorithm}{Algorithm}
\usepackage{graphics}
\usepackage{graphicx}
\usepackage{epstopdf}
\usepackage{ed}
\usepackage{wrapfig}
\usepackage{subfig}

\usepackage{booktabs}
\newcommand{\ra}[1]{\renewcommand{\arraystretch}{#1}}

\usepackage{rotating}
\usepackage{epsfig}
\usepackage{epsf}
\input{epsf}

\usepackage{booktabs}
\newcommand*{\THESIS}{}%

\makeatletter
\renewcommand\section{\@startsection{section}{1}{\z@}%
                       {5\p@ \@plus 2\p@ \@minus 2\p@}%
                       {5\p@ \@plus 2\p@ \@minus 2\p@}%
                       {\large \bfseries \raggedright}}
\renewcommand\subsection{\@startsection{subsection}{2}{\z@}
{5\p@ \@plus 4\p@ \@minus 4\p@}%
{3\p@ \@plus 4\p@ \@minus 4\p@}%
{\normalsize \bfseries \raggedright}}
\makeatother
\usepackage{url}

\makeatother

\begin{document}
\title{LFTL: A multi-threaded FTL for a Parallel IO Flash Card under Linux}



\author{%
{Srimugunthan{\small $~^{1}$} , K. Gopinath{\small $~^{2}$}, Giridhar Appaji Nag Yasa{\small $~^{3}$}  }%
\vspace{1.6mm}\\
\fontsize{10}{10}\selectfont
  \begin{center} \hspace{2mm}Indian Institute of Science$^{[}$$~^{1}$$~^{2}$$^{]}$ \hspace{10mm} NetApp$~^{3}$\\ \end{center}
                                                                                
 \begin{center} \hspace{2mm}Bangalore                            \hspace{16mm} \  \\ \end{center}

\fontsize{9}{9}\selectfont\ttfamily\upshape
$~^{1}$srimugunth@csa.iisc.ernet.in 
$~^{2}$gopi@csa.iisc.ernet.in 
$~^{3}$giridhar@netapp.com

}

\maketitle 


\maketitle


\begin{abstract}
\textbf{
     New PCI-e flash cards and SSDs supporting  over 100,000 IOPs are now available,
  with several usecases in the design of a high performance storage system. By using an
  array of flash chips, arranged in multiple banks, large capacities   
  are achieved. Such multi-banked architecture allow parallel read, write
  and erase operations. In a raw PCI-e  flash card, such  parallelism is directly 
  available to the software layer. In addition, the devices have restrictions 
  such as, pages within a block can only be written sequentially. The devices also have 
  larger minimum   write sizes ($>$4KB).  Current flash translation layers (FTLs) in Linux are not well suited
  for such devices due to the high device speeds, architectural  restrictions 
  as well as  other factors such as high lock contention.  We present a FTL for
  Linux that takes  into account the  hardware restrictions, that also exploits 
  the parallelism to achieve high speeds. We also consider leveraging the 
  parallelism for garbage collection by scheduling the garbage collection activities on 
  idle banks. We propose and evaluate an adaptive method to vary the amount of garbage collection
  according to the current I/O load on the device.
   
}

\end{abstract}

\section{Introduction}

Flash memory technologies include NAND flash, NOR flash, SLC/MLC flash
memories, and their hybrids.  Flash memory technologies, due to the
cost, speed and power characteristics, are used at different levels
of the memory hierarchy.

  Flash memory chips were dominantly used in embedded systems for handheld  devices.
 With recent advances, Flash has graduated from being only a low performance consumer
technology to also being used in the high performance enterprise
area. Presently, Flash memory chips are used as extended system memory, or as a PCI 
express card acting as a cache in a disk based storage system, or as a SSD drive completely 
 substituting disks.

  However, flash memory has some peculiarities such as that a write is
possible only on a fully erased block. Furthermore, a block can be erased
and rewritten only a limited number of times. Currently, the
granularity of an erase (in terms of blocks) is much bigger than the
write or read (in terms of pages). As the reuse of blocks depends on
the lifetimes of data, different blocks may be rewritten different
number of times; hence, to ensure reliability special techniques called 
wearlevelling  are used to distribute the writes evenly across all the blocks.

 Multiple terabyte-sized SSD block devices made with multiple flash chips are presently available.
New PCI-e flash cards and SSDs supporting high IOPS rate
(eg. greater than 100,000) are also available. Due to limitations in scaling of the size of flash memory 
per chip, parallelism is an inherent feature in all of these large flash storage devices.

\textbf{\textit{Flash memory scaling}}

 IC fabrication processes are characterized by a feature size, the
 minimum size that can be marked reliably during manufacturing. The feature size determines the 
 surface area of a transistor, and hence the transistor count per unit area of silicon.
 As the feature size decreases, the storage capacity of a flash memory chip increases.
 But as we reduce the process feature size, there are problems like, the bits can not be stored 
 reliably. Sub-32 nm flash memory scaling has been a  challenge in the past. Process innovations
 and scaling by storing multiple bits per cell have enabled bigger flash sizes per chip. 
 As of this writing, the minimum feature size achieved is 19nm, resulting in an 8GB flash chip storing 2 bits per
 cell. But storing multiple cells per chip adversely affect the endurance of the flash\cite{anandtech} bringing
 down the maximum erase cycles per block in the range of 5000 to 10,000.  It has been shown that scaling by packing more bits per
 cell degrades the write performance\cite{flashbleak}. Also the bus interface from a  single NAND chip is usually
 a slow asynchronous interface operating at 40MHz.   So in order to achieve larger capacities and faster speeds,
 multiple flash memory chips are arrayed together and they are accessed simultaneously. These kind of architectures 
are presently used in SSDs\cite{DToff08}

\textbf{\textit{SSD architecture and interface}}
    An SSD package internally  has its own memory (usually battery backed DRAM
    ), a flash controller and some firmware that implements a
    flash translation layer which does some basic wearlevelling and
    exposes a block interface. Hard disks can be readily substituted
    by SSDs, because the internal architecture of SSDs is hidden behind
    the block interface. Because of the blackbox nature of SSDs, some
    changes have been proposed in the ATA protocol interface such as
    the new TRIM command, that marks a block as garbage as a hint
    from an upper layer such as a filesystem to the SSDs\cite{TRIM}.

\textbf{\textit{Raw flash card vs SSDs}}

Though most flash is sold in the form of SSDs, there are also a few
        raw flash cards used in some scenarios. Such flash cards 
pack hundreds of gigabytes in a single PCI-e card with an SSD-like
architecture but without  any
on-board RAM or any firmware for wearlevelling.

	The flash card that we studied is made of several channels or interfaces. Several of 
  flash chips make a bank and several of a banks make a channel or interface. It is possible to do
  writes or reads simultaneously across banks.  The flash card have an in-built FPGA that supports DMA
   write size  of 32KB and DMA read size of 4KB.  Pages within a block can only be written sequentially
   one after another.

  It is important to mention that we have used a raw flash card and not a SSD for our study.

\textbf{\textit{Flash translation layer}}

  The management of raw flash can either be implemented within a  device driver or other software layers above it.
  The layer of software between the filesystem and the device driver that does the flash management is called Flash
   translation layer (FTL). The main functions of the FTL are address mapping, wearlevelling and garbage collection.
   Working in the context of Linux, we have  designed a flash translation layer for a raw flash card which exploits 
  the parallel IO capabilities. In addition to the main functionalities our FTL also does caching of data, before it 
  writes to the flash.

The following are the main contributions of this work

\begin{itemize}
 \item We present a design of a Flash translation layer(FTL) under Linux, which can scale to higher speeds.
 Our FTL also copes up with device limitations  like larger minimum write sizes by making use of buffering inside the FTL. 
 \item We exploit the parallelism of a Flash card with respect to block allocation, garbage collection and for initial device scanning. 
  The initial scan is  time consuming for larger flash; hence exploiting parallelism here is useful.
\item  We give a adaptive method for varying the amount of on-going garbage collection according to the current I/O load.
\end{itemize}

      Section 2 presents the background with respect to flash memories, flash filesystems, FTLs and the flash card used.
    Section 3 describes the design of our FTL.
    Section 4 presents the results. 
    Section 5 is a discussion of some issues in our FTL and also about future directions of work
    and Section 6 concludes.

\section{ Background }

  Flash memory has a limited budget of erase cycles per block. For SLC
 flash memories it is in hundreds of thousands while for MLC flash it is in
 tens of thousands.
 
  Wearlevelling is a technique, that ensures that all the blocks utilise
their erase cycle budget at about the same rate. Associated with wearlevelling is
garbage collection. Wearlevelling and garbage collection have three components:
\begin{itemize}
 \item Out of place updates (referred  to as \textit{Dynamic
   wearlevelling}): A rewrite of a page is written to a different page on the flash. The old
    page is marked as invalid. 
 \item To reclaim invalidated pages (referred to as \textit{garbage
   collection}): To make the invalid pages writable again, the block
   with invalid pages has to erased. If the block has some valid
   pages, it has to be copied to  another block.
 \item Move older unmodified data to other blocks of similar lifetimes (referred to
  as \textit{Static wearlevelling} )
\end{itemize} 

  Because of the out-of-place update nature of the flash media, there is
 a difference between a logical address that a filesystem uses and a
 physical address that is used for writing to the flash. This logical address
 to physical address translation is provided by the flash translation layer.

 Note that both garbage collection and static wearlevelling incur some extra writes.
 The metric, write amplification, quantifies the overhead due to garbage collection 
 and static wearlevelling. Write amplification is the ratio between the number of user
  writes and the total writes that happened on the flash. A good wearlevelling algorithm 
  minimises the write amplification as much as possible.

  When Garbage collection and static wearlevelling are done during idle time when there
  is no I/O happening on the system, the overhead is only the extra writes and erases. 
  But, under a heavy, continuous IO load,  when the amount of free space drops low, it
  is necessary to invoke garbage collection as part of a user write, to create a fresh 
  supply of free blocks. In this case garbage collection  increases  the latency of every
  write request. A policy decides when the garbage collection is to be triggered and how 
  much space should be freed in one invocation.

  Based on these goals several interesting algorithms have been proposed. The 
  reference\cite{Gal05} is an excellent survey of flash memory related algorithms.\newline

\subsection{FTLs}


Unlike SSDs, the FTL has to be implemented in software for raw flash cards. 
The FTL exposes a block device interface to filesystem and hence any disk file system
can be mounted  on the raw flash card. The FTL does out-of-place updates on the flash by  
maintaining a table of logical to physical address translation entries. 
The translation entries can be in units of pages, storing logical page number to physical 
page number. For a large flash of size 1TB and 4KB page size, assuming 4 bytes per translation
 entry we need 1 GB of space for the mapping table. Instead of storing the the translation table 
 entries in the granularity of pages, we can store in the granularity of blocks storing logical block
 number to physical block number. The disadvantage of this approach is that even if only 
a single page in a block needs to be modified, we have to re-write all the other pages to a new block.

So instead of  a pure page based or block based mapping tables, popular FTLs use modified mapping procedures.
 Some well known FTLs are NFTL\cite{NFTL}, FAST FTL\cite{FASTftl},  and DFTL\cite{DFTL}.


 NFTL is one of the Linux FTLs and is available as part of the linux code base.
 In Linux, flash devices are treated specially as \textit{mtd} (memory technology devices) devices.
 The flash specific characteristics like the asymmetric sizes and asymmetric access speeds for read, 
 write and erase, the  presence of bad blocks in flash, reading or writing to the spare area of the 
 flash, makes it different from a block device.  The NFTL operates over a mtd device and exposes a 
 block device, with  sector size of 512 bytes, to any filesystem above it.


  In NFTL, a
  few blocks are designated as replacement blocks. The remaining
  blocks are called primary blocks. Any re-write of a page in a primary block is written to a replacement
 block, with the same page offset. If there is one more re-write of the same page,
  another replacement block is allocated  and the 
 corresponding page is written. So there is  chain of replacement blocks that are  allocated to a primary block.
 The spare area of a replacement block is used to point to the next replacement block in the chain.
  The mapping table that is in memory stores only the
  mapping of a logical block address to physical address of the primary block.
  From the primary block, the chain of replacement blocks should be traversed to find
  the latest page.  When availability of free blocks drop low, garbage collection, which is also
 called a fold operation, selects the longest chain and merges the up-to-date 
 pages to one block.  The NFTL algorithm will not work for flash devices that 
 require sequential writing of pages, as the pages in replacement block are
 written non-sequentially.

      The FAST FTL\cite{FASTftl} is similar to NFTL but the
      association between pages in the replacement block (also called
      as log blocks) and the primary block is different. Here there is
      no strict association between the replacement block and primary
      block. A replacement block can contain pages of any primary
      block. It is also not necessary that re-writes of a page are written to the 
     same page offset, within   the replacement block. When availability of replacement
     blocks drop low, one of the replacement block has to be freed, after  pages in it
     are written to primary blocks. This is called the merge operation. Depending on the
     number of valid up-to-date pages in a replacement block, the merge cost varies. 
     This merge cost is smaller for a sequential IO workload. For random IO, the merge cost
    can be significant.

  NFTL and FAST FTL come into the category of hybrid FTL, where there is a  mapping table at a block 
level granularity for all blocks which is stored in RAM. The page level translation is  done through
 information stored on the flash. The in memory page level tables don't store the translation entries
 for all the pages, but only for a subset of pages.

	Instead of using replacement block scheme and hybrid translation, the DFTL  implements a page based FTL,
   but the page table is maintained both in SRAM and on the flash. The SRAM stores most recently 
  accessed page table entries. By not following the replacement block scheme, DFTL avoids 
  the merge operation, that needs to be done, when the replacement blocks are merged with
   their primary blocks. The blocks which store the map table entries are designated as 
   translation pages/blocks. The translation blocks exist alongside the normal data blocks. A 
  in-memory global translation directory stores the location of the translation pages. When a 
  SRAM entry is evicted, the page table entry has to be written to translation page on flash. 
  It is proposed to use a combination of batched and lazy updates of translation entries.
  Because  the map table is stored on flash, when a data block is garbage collected, all the translation pages
 in flash have to be updated, to reflect the new data page locations.  The presence of temporal locality
 in the workload helps the DFTL algorithm, by facilitating the SRAM hits and the lazy and batched updates of
  translation table entries.
  
  Almost all SSDs come with some amount of RAM to absorb random writes. The buffer management schemes
  BPLRU \cite{BPLRU} and PUDLRU\cite{PUDLRU} have studied the influence of using caches in FTLs. The 
  size of the caches and the replacement policy for the buffers are the main points of study in these papers.

  The BPLRU scheme\cite{BPLRU} assumes a hybrid FTL and  uses RAM cache  to reduce the number of erases.
  A LRU list is maintained in the granularity of a flash block. If any sector within the block is accessed, 
  the recency position of its corresponding block is modified in the LRU list.    To reduce merge costs of hybrid FTL,
  the pages in a block are pre-read (called page-padding) and kept in the cache.  The PUDLRU scheme \cite{PUDLRU}  
  has extended the BPLRU replacement policy by taking into account both frequency and recency of access. They also
  show that buffer management is essential for any FTL.

\ifdefined\THESIS
\else

\begin{figure*}[]
\begin{center}
 \includegraphics[scale=0.2]{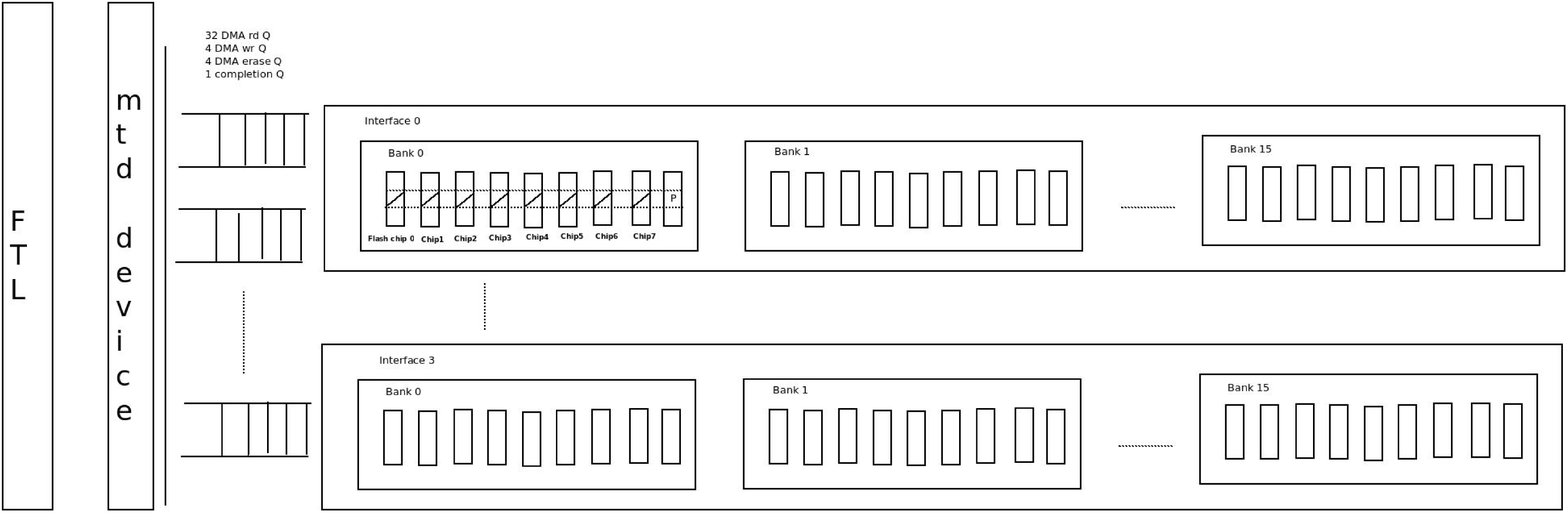}
\caption{Flash card hardware architecture}
\label{fig:virgoarch}
\end{center}
\end{figure*}   

\fi

\subsection{Flash file systems}

  There are flash specific filesystems like yaffs2, logfs and UBIFS that implement some or all 
 of the FTL functionalities like wearlevelling and garbage collection in a filesystem.

  Yaffs2 \cite{yaffs2howto} has to be mounted on a mtdblock device. The mtdblock  driver
 accumulates the writes to a physical block in a buffer and writes to the flash when it is full.  

  Yaffs2 is a log structured file system, where the the whole flash storage is seen as a log.
 The data is sequentially written in the flash media, irrespective of a write or a re-write.
 The block allocation algorithm selects the next block to write as current block written plus one, 
 if it is available. If the next block is not available, we again search sequentially for the next 
 free block. There is a sequence number that is written along with every write to order the log in 
 chronological order.

  Garbage collection algorithm searches for a block with very few valid pages as a `victim' block
 for garbage collection.  There are many levels of garbage collection, like background garbage collection,
  passive garbage collection and aggressive garbage collection, which are triggered depending on the number 
  of free blocks available. These levels differ by the condition when they are invoked, the threshold for number
  of dirty pages for victim selection and the number of blocks to be searched for a victim.  When the number of
  free blocks is sufficient, only background garbage collection runs. The background garbage collector runs as a
  separate kernel thread and it wakes itself up after regular intervals and does garbage collection, if necessary.
  Passive and aggressive garbage collection happen before a filesystem write request is serviced. For passive and 
  background garbage collection, the search window is  limited to last victim block plus  atmost  100 blocks. For
  aggressive garbage collection, the entire flash is searched for a victim garbage collection block. Because 
  aggressive garbage collection is invoked when free space is running very low,  blocks with fewer dirty pages are 
  also selected as victim blocks.

  Yaffs2 has checkpointing feature which stores all the filesystem information, present in memory to the flash
  media during unmount. On the subsequent mount, the RAM structures are reconstructed using the checkpoint. If there was an improper
  unmount, due to power failure, yaffs2 does a scan of the media, and  using the sequence numbers,  reconstructs 
  the filesystem structures,

   UBIFS is different from yaffs2 and logfs, in that, it operates over UBI layer. The UBI layer is a block mapping
   layer which exposes a set of logical erase blocks to the UBIFS. When the UBI module is loaded,  the block mapping
   information is read from the flash.  UBIFS maintains the filesystem tree on flash. A leaf update of the tree, because
   of out-of-place update, also requires that we update all its parent nodes. This is called as the wandering tree problem. 
   UBIFS uses a tree node cache and journalling to reduce the amount of on-flash updates needed for tree-index updates.  
   When a logical erase block is garbage collected, the blocks which store the tree-indexes have to be treated differently.  
   So, it is  UBIFS and not the UBI, that decides when a logical erase block can be erased completely.  In a multi banked flash
   card, as  UBIFS operates only on logical erase blocks,  it is not straight forward to parallelise the garbage collection 
   activity, to an idle bank,  UBIFS has been a successful filesystem in the embedded scenario because it tolerates power
   failures very well.

  Logfs is a new flash filesystem  that can operate directly over an mtd device.  Logfs like UBIFS maintains the filesystem 
 tree on flash and also faces the wandering tree problem.  Logfs mitigates the on-flash updates, by using ``aliases''  which
 remembers that we have done an update to a leaf node. The updates to the parent nodes are delayed. The aliases are stored as
 part of a on-flash journal. Logfs actually has two journals. When the second journal is written heavily, for wearlevelling
 purposes, its position on the flash is changed. The new position of the second journal is maintained in the first journal.
 The aliases are recorded in the second journal.  In logfs, there is no separate area in the on-disk structure for inodes.
 All the inodes are stored in a separate file called the inode file. The use of inode file make the inode metadata writes, 
 same as log structured file writes.   The inode for the inode file (called master inode) is stored as part of the second journal.  
 The mount time in logfs is only the time taken to  replay the journal.

 Logfs was accepted in the kernel 2.6.34 version, but the development pace of logfs has been relatively slow.
As Logfs has not attained full stability, we present
 comparative evaluation with yaffs2 and UBIFS only.  

We next discuss the Flash card that we used.

\subsection{Flash Card Architecture}

The flash card, studied  is of the type that is  used for I/O acceleration,
 by caching recently used user data and metadata in a SAN/NAS storage system .

  The flash card  is of size 512GB. 

  The Flash card is made of 4 interfaces or channels, 16 banks per interface and 8 data
 chips + 1 parity chip per bank. Chips are SLC  nand chips  with 1GB size and a maximum
 of 100,000 erase cycles as endurance limit.

  The Card has a built-in FPGA controller that provides DMA channels for reads, writes and erases. 
  A single DMA write size is 32 KB that is  striped  equally among all 8 chips in a bank with the 
  parity generated  in the ninth chip. The DMA read size is 4KB and it reads the striped data across 
  8 chips. DMA erase makes 2MB of data across 8 chips to become all ones. For every 512 bytes, BCH 
  checksum of 8 bytes is generated by hardware. An extra 8 bytes per 512 bytes is available for user
  spare data. So  there is as much as 512 spare bytes  per 32KB page available for storing any user
  metadata. The device has programmable registers through PCI interface. We have written a driver for
  this device that exposes a 512GB mtd device and with  a write size of 32KB and an erase size of 2MB.

      The device offers parallel read, write and erase support across banks. There are 32  DMA read
   queues, 4 DMA write queues and 4 DMA erase queues. For write and erase queues, the device driver 
   assigns each of the four queues  to each of the 4 interfaces and queues requests correspondingly.

\ifdefined\THESIS

\begin{figure*}[htbp]
\begin{center}
 \includegraphics[scale=0.2]{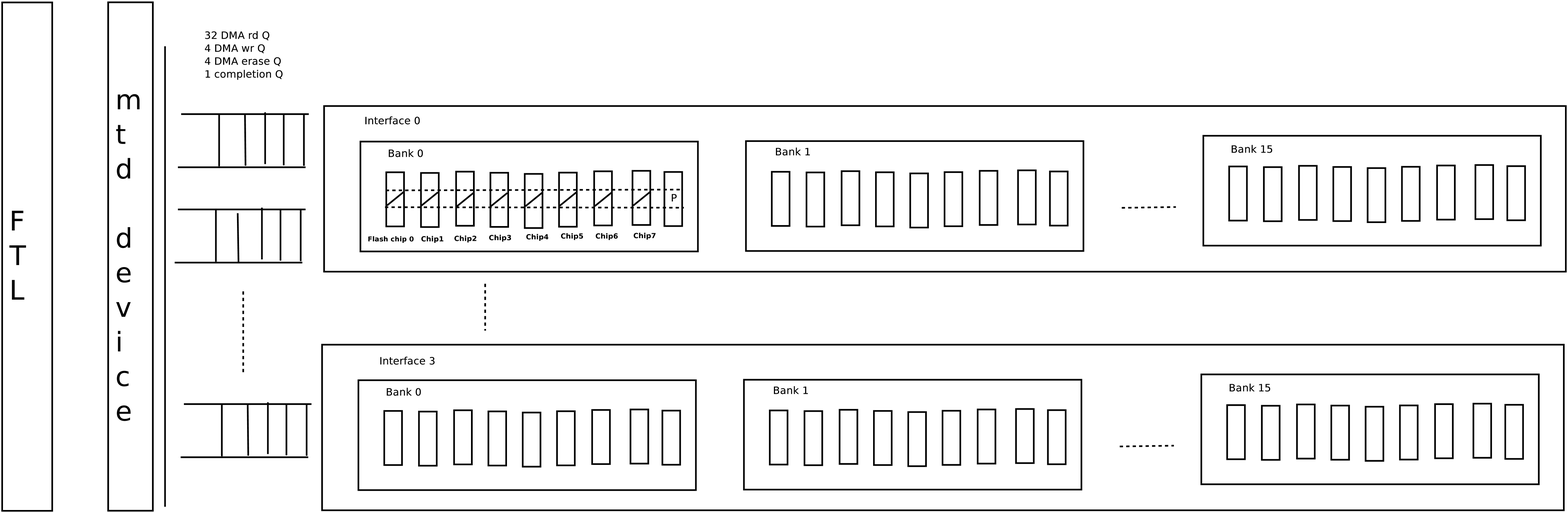}
\caption{Flash card hardware architecture}
\label{fig:virgoarch}
\end{center}
\end{figure*}

\fi

 The hardware FPGA controller services these queues in round robin fashion.  For read queue, the 
  device driver assigns 8 queues per interface. Each queue is assigned for 2 successive banks of an interface. 
   Each queue can hold 256 requests. When a request is completed, the device responds by writing a
  completion descriptor in a completion queue. The completion queue is of size 1024 entries. When requests
   proceed in parallel, it is possible to receive completion descriptors in out of order fashion. 

     The driver also has a rudimentary bad block management.  Bad blocks are identified initially
    and stored in an on-disk file in the host system. The driver during initialisation reads from this file 
    and constructs its in memory bad block table.

\subsection{Parallelism in the flash device}

A flash based SSD has many levels of parallelism. 

  Firstly, there are many channels. Each channel is made of flash
 packages, each of which consists of several flash chips. Within a
 single flash chip, there are multiple dies and finally there are
 several flash planes within a single die. The flash chips are mostly
 connected using a 40MHz bus in our device. The operating  bus speed
 increases, as we move from inner flash chips to outer flash packages.
 The flash chips can be simultaneously read to internal
 buffers. The transfer from these internal buffers happen at faster bus
 speeds. Because of the hierarchical architecture of channels, flash
 packages, chips, dies and planes, there are several levels of
 parallelism available. 

  At the lowest level, a feature  called \textit{multiplane
  command} allows 2 operations of the same type (read, write or
  erase) on planes within a die to happen concurrently. Two dies
  within a chip share the IO bus in an interleaved
  fashion. \textit{Interleaving} command executes several page read,
  page write, block erase and multi-plane read/write/erase operations in
  different dies of the same chip in a pipelined fashion.  At the highest
  level, two  requests targeted to 2 different channels or flash packages can 
  execute simultaneously
  and independently.

  Although the parallelism is known to NAND controller inside a SSD, the amount
  of parallelism that is exposed to software directly is very limited. Usually,
  SSD parallelism is exploited  indirectly, like writing in larger sizes.

\begin{table}[h]

   \begin{small}
    \begin{tabular}{|l |l |l | l |}
\hline    

\textit{\textbf{Banks }}  	& {\textbf{Num   }}   & {\textbf{Read   }}   & {\textbf{Write   }}  \\  
       {\textbf{  used }} 	& {\textbf{pthreads }} 	& {\textbf{speed }} & {\textbf{speed }} \\
   \hline
   \hline
    Bank0  only 		 &   16 		&   114 MB/sec 	&   82 MB/sec 	 \\ \hline
    Bank0 and 1	  		 &   16 		&   114 MB/sec 	&   155 MB/sec 	 \\ \hline
    Bank0 and 2	  		 &   16 		&   228 MB/sec 	&   160 MB/sec 	 \\ \hline
    Bank0 and 32  		 &   16 		&   228 MB/sec 	&   160 MB/sec 	 \\ \hline
    Bank0,1,2,3 		 &   16 		&   228 MB/sec 	&   246 MB/sec 	 \\ \hline
    Bank0,16,32,48 		 &   16 		&   320 MB/sec 	&   320 MB/sec 	 \\ \hline
    Bank0 to 15	 		 &   16 		&   290 MB/sec 	&   246 MB/sec 	 \\ \hline
    Bank0 to 32	 		 &   32 		&   304 MB/sec 	&   492 MB/sec 	 \\ \hline
    Bank0 to 64	 		 &   32 		&   304 MB/sec \footnotemark	&   533 MB/sec 	 \\ \hline

    \end{tabular}

    \end{small} 
\captionsetup{singlelinecheck=off,justification=raggedright}
\caption{Measurements directly over the device driver}  
\label{tab:driverspeed}

\end{table} 

\footnotetext[1]{Currently it is not  clear  why the   raw read speed  doesn't scale as much as write speed.}

 In the flash card that we used, there is a  hierarchy of channels,
 banks and chips. The hardware FPGA employs DMA read or write that
 stripes the data within the chips. After the driver enqueues the
 read/write/erase requests in the DMA queues, the hardware FPGA
 services these queues in a Round-robin fashion. So the exposed
 parallelism is at the bank and channel levels.

    The table \ref{tab:driverspeed} shows the driver level measurement,
    taken by reading or writing directly  the mtd device from several
    pthreads. Each pthread reads or writes 200MB of data in 32K
    chunks. The read and write bandwidth increases as the banks are
    simultaneously utilised.  The queue to bank assignment also
    affects the achieved bandwidth. For example, the read bandwidth
    when bank0 and 1 are used, is lesser than that for bank0 and 2. This
    is because bank 0 and 1 share the same DMA queue while bank 0 and
    2 do not.

  We can see that compared to using banks within the same channel, when banks across
 channels are utilised, we can achieve greater speed. 

  For simplicity and generality, we assume only  bank level parallelism in flash card 
 and try to exploit this from a flash translation layer.

\FloatBarrier
\section{ Design of the Flash Translation Layer}

 We first discuss the performance bottlenecks in the present design of linux FTLs.
 We then discuss the design we adopted and  write and read algorithms along with 
 buffer management to make use of parallelism of the flash card. We further exploit 
 the parallelism for garbage collection activity.

\subsection{Performance Bottleneck in Linux FTLs}

Current implementations of FTL in linux, like NFTL and mtdblock  use the mtd\_blkdevs infrastructure.
 The mtd\_blkdevs basically registers for a new block device and a corresponding request processing
 function.  This mtd\_blkdevs infrastructure is used by any FTL which have their own implementations for 
 reading and writing sectors from the device.

\ifdefined\THESIS

\begin{figure}[h]
\begin{center}

 \includegraphics[scale = 0.4] {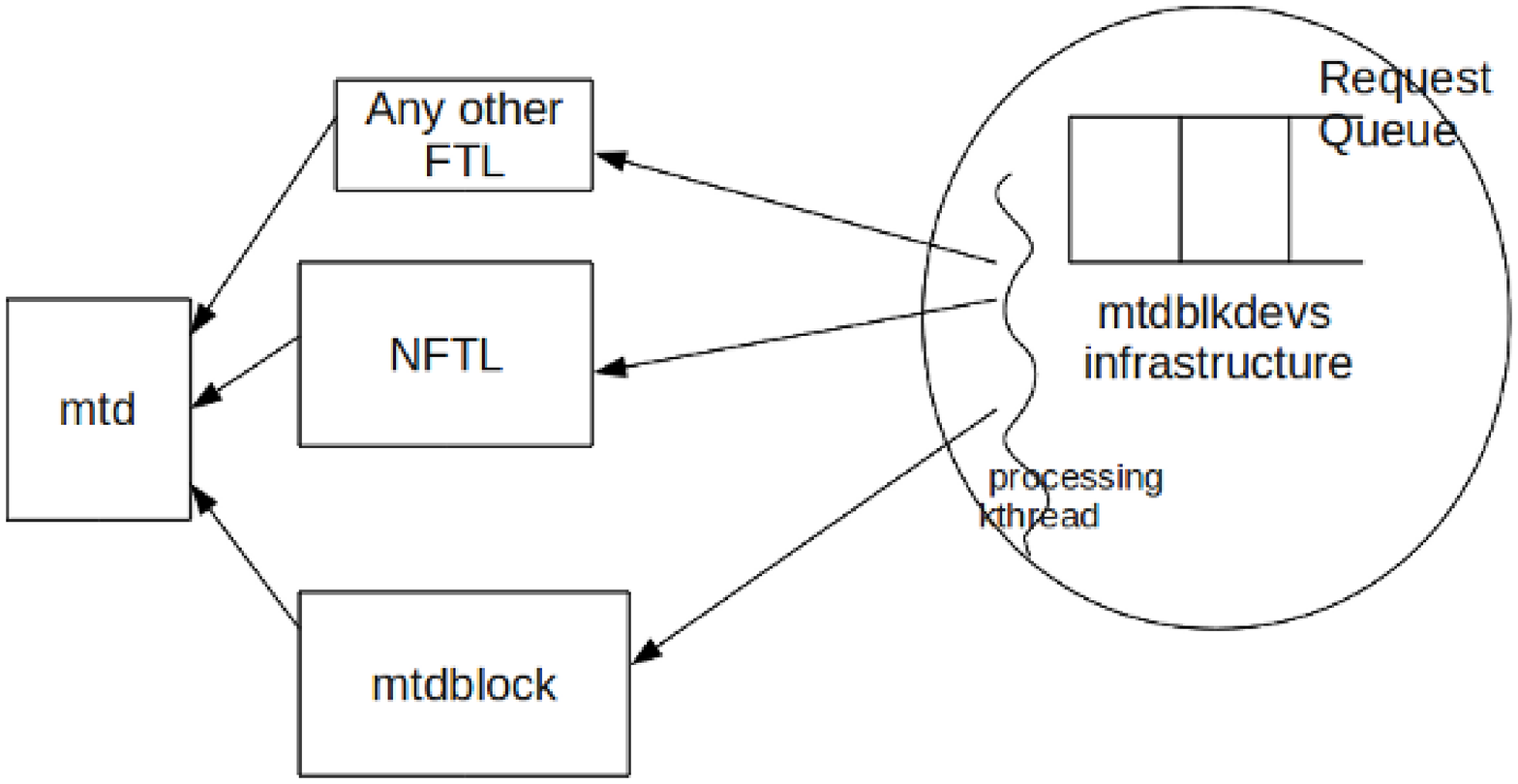}
 
\caption{ mtdblkdevs infrastructure in Linux}
\label{fig:linuxFTL}

\end{center}
\end{figure}

\fi

 It is to be noted that the NFTL and mtdblock are pseudo block devices. In Linux a  block device's request processing
 function  is not supposed to sleep. However, the threads that call mtd apis can sleep, and so the mtd\_blkdevs does the request 
 processing from a separate kthread. As shown in Figure \ref{fig:linuxFTL}, the kernel thread receives the IO requests from 
 the block layer's request  queue and calls NFTL or mtdblock's write/read sector code.  The actual reading/writing of the
  IO request is done  by the NFTL or mtdblock's write or read sector code. The kernel threads sleep when the request queue is empty.
  When  a new request is  enqueued the mtd\_blkdevs request   processing function wakes up the kernel thread, which does the actual  processing.

  The block layer provides one request queue for a registered device. The block layer's request queue is used by the
 IO scheduling algorithms.  The request queue is a doubly linked list and it  is used for  coalescing and merging 
 the bios (block IO requests) to much larger requests. The IO scheduling algorithms of the block layer are specifically targeted for hard 
 disks to minimise seek time. Flash memories don't benefit from the IO scheduling optimisations of the block layer. The 
 merging and coalescing optimisations also are not useful because, our FTL uses a list of buffers that are used to accumulate 
 the IOs to larger sizes. In addition to those reasons,  using a single request queue, means that we have to take a lock 
 for dequeueing the IO request. The Figure \ref{fig:mtdblkdevs} is the code snippet that is implemented in mtd\_blkdevs.c 
 and it shows the acquiring and releasing of request queue locks.  The problem of request queue lock contention as a hindrance to  high IOPS
 devices has been discussed in the linux community \cite{lwnmq}. These reasons motivated us to not use the block layer's request
  queue management.

 \ifdefined\THESIS

\begin{figure}[t]
\begin{center}
 \includegraphics[scale = 0.35] {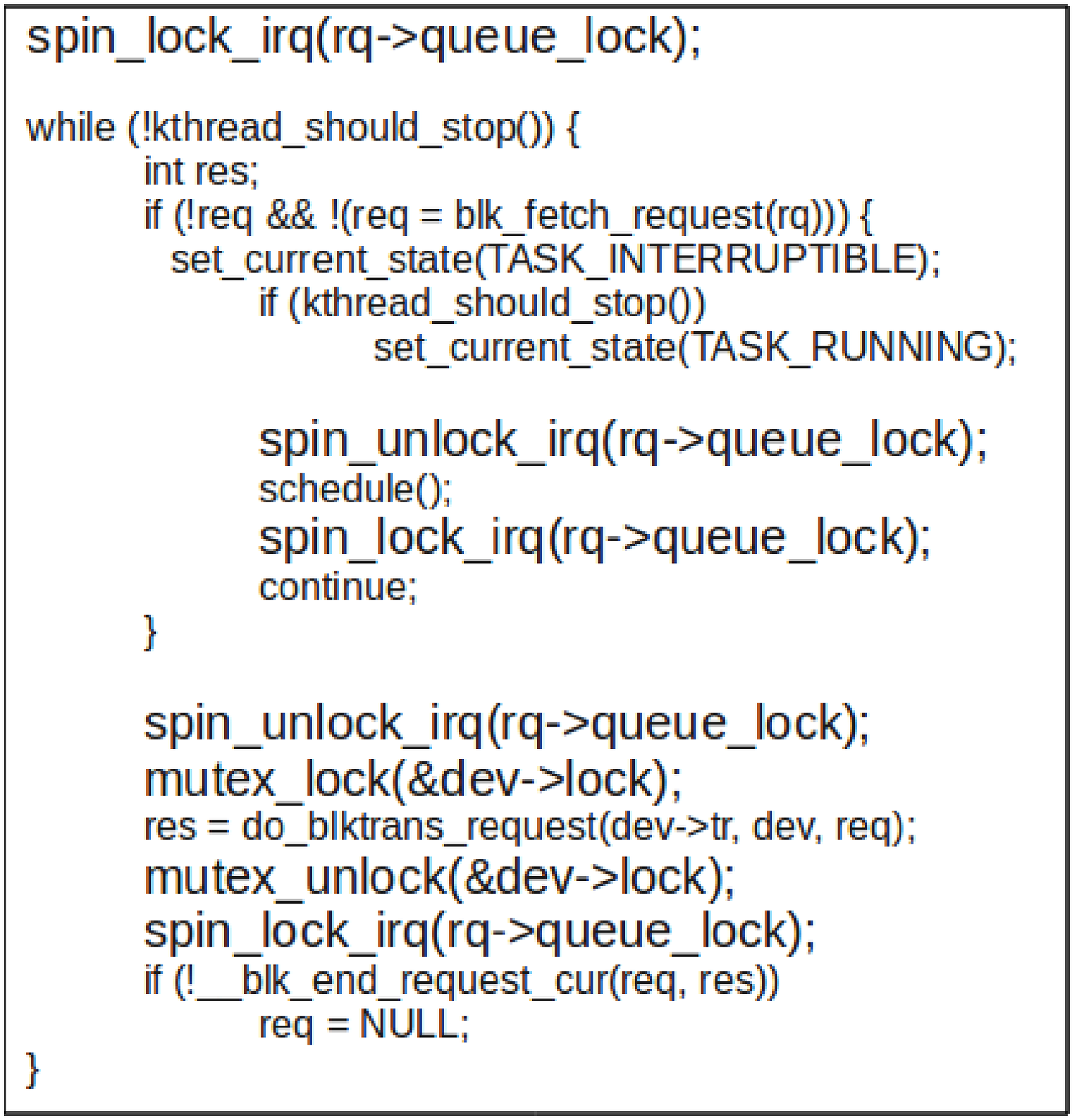}

\caption{ mtdblkdevs code showing lock contention}
\label{fig:mtdblkdevs}

\end{center}
\end{figure}
\FloatBarrier

\fi

  In the code snippet of Figure \ref{fig:mtdblkdevs} a global device lock (dev-$>$lock) is used assuming 
 that requests to the flash device can only be done serially one after another.  With the existing mtdblkdevs
 infrastructure, the maximum IO bandwidth that we were able to obtain was only upto 55MB/sec. The large capacity flash 
 cards can process several IO requests simultaneously on several banks. Since banks are the minimal unit of
  parallel IO, the global device level locks should be split in to per bank locks.

  We incorporated these changes in the design of our FTL and Figure \ref{fig:FTLflow} shows the design that we 
 followed. In our FTL, We  maintain multiple FIFO queues of our own and each of these queues has an associated 
 kernel thread. The kernel  threads  processes I/O requests from its corresponding queue. So these multiple
 `FTL I/O kernel threads' provide  parallel I/O capability.

    The data structure corresponding to block layer's I/O request is called \textit{bio}. The entry point in the
  linux block I/O layer is the  function  bio\_submit() which receives the bio. We intercept the bios directly 
  and direct the bios to one of our FIFO queues. The intercepting of the bios is also done for RAID devices and
  the kernel exports an API for registering our custom \textit{make\_request} function.  

  \ifdefined\THESIS
\begin{figure*}[htb]

\centering
\includegraphics[scale=0.35]{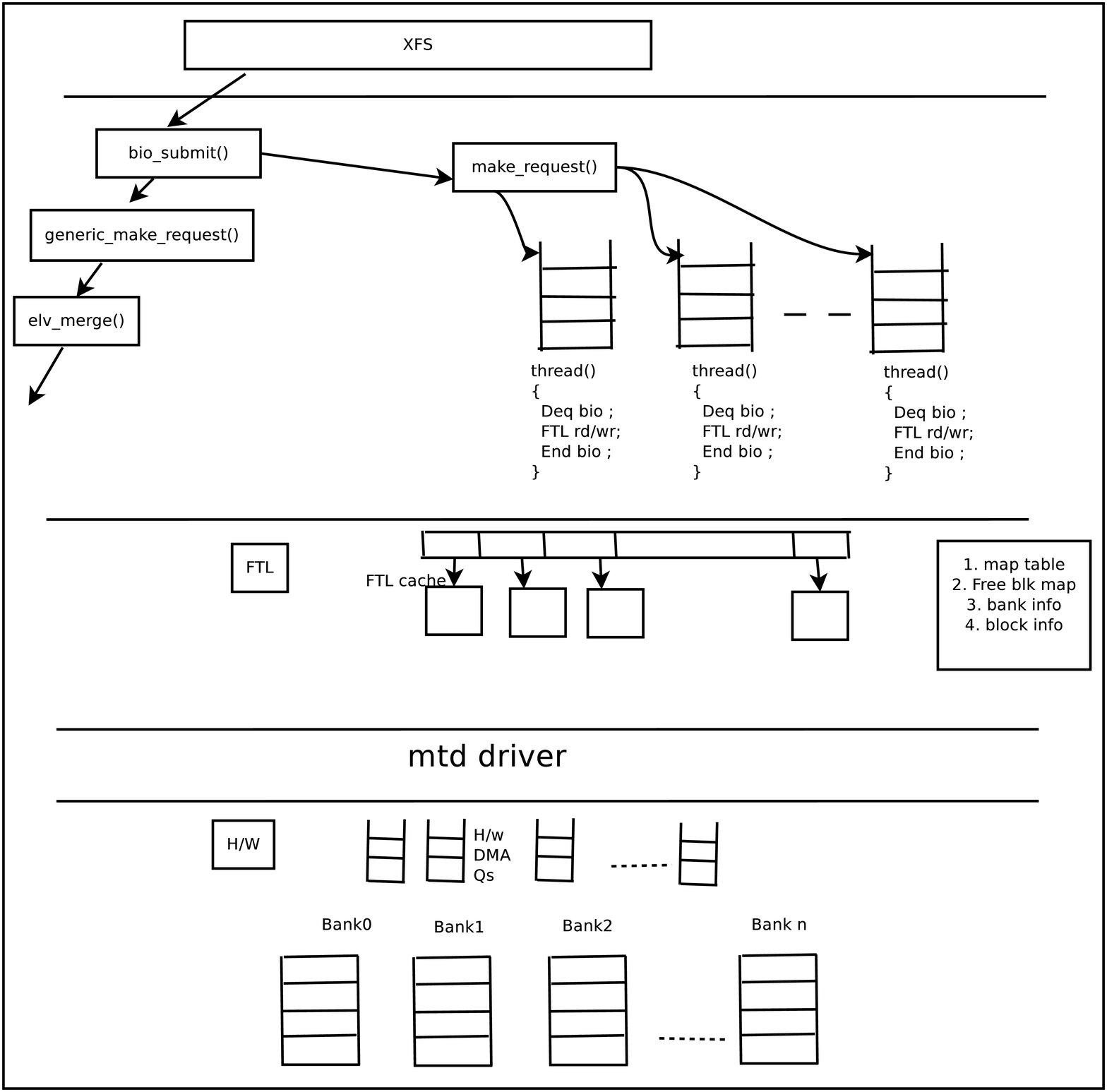}
\caption{FTL design}
\label{fig:FTLflow}
\end{figure*}
\fi

    The mtd API for read and write are synchronous API calls. The use of multiple FIFO queues and  multiple 
 kernel threads, in addition to removing the lock contention, also facilitates several IOs to be in-flight 
 simultaneously, even though the mtd read and write calls are synchronous. The per-queue kernel thread calls 
 the read/write sector functions of our FTL.

    Unlike NFTL we have  exposed a 4KB block device size, but the minimum flash writesize is 32KB. So FTL uses 
 a set of buffers and accumulates the eight consecutive logical 4K data to a single 32K buffer and finally writes
 it to the flash. The per-buffer size and the number of buffers in the FTL are module parameters. The per-buffer size 
 should be atleast the flash page size. The number of buffers is only limited by the amount of RAM in the host system. 
 Currently we have used 256 buffers and the buffer size is 32K.

\subsection{Write and Read algorithm}

The main data structures, that are used in our FTL, are listed below.

\begin{itemize}
\item \textit{FreeBlksBitMap}:  The FreeBlksBitMap indicates whether a given block is free or occupied. 
 The Freeblks Bit-map datastructure is protected by  per-bank locks.
\item \textit{BufferLookUp Table}: The BufferLookup Table stores the the logical page number that is stored 
 in a buffer. There is no lock taken for reading or writing the buffer Lookup table.
\item \textit{BlkInfo}: The blkinfo stores  a bitmap indicating the valid pages in the block and the number 
 of valid pages in the block.
\item \textit{BankInfo}:  The bankinfo stores the number of free blocks and number of valid pages in the bank. 
 We used linux atomic bitmap operations and atomic variable updates for modifying blkinfo and bankinfo.
\item \textit{MapTable}:  The maptable stores 4 bytes/entry for every page in the flash.  For 512GB flash and 
 32K page size, the map table is of 64MB size. We have used bit-locking, with a single bit protecting  per-entry
 of the  maptable. We used the linux atomic bitmap operations for the bit-locking. The space for bitmap need not 
 be allocated  separately, as we can steal one  bit from the map table entry.
\item \textit{AllocBufBitmap}: The AllocBufBitmap  indicates that we are trying to allocate a buffer for a 
 logical page. It is used for mutual exclusion to prevent allocation of two buffers to the same logical page number. 
\end{itemize}

    The write algorithm is given in Algorithm 1.  Initially we search the buffer lookup table, if the given \textit{lpn}
  (logical page number) is  present in any of the FTL cache buffers. If the lpn is available, we take the per 
  buffer lock and again check if the buffer corresponds to lpn. This is similar to double-checked locking
  programming idiom\cite{doublecheck}.  If the lpn is not in the FTL cache, then we have to  allocate one of 
 the buffers to this lpn. We set  the bit in AllocBufBitmap  for this lpn to indicate that we are trying to 
 allocate a buffer for this lpn. After taking the bit-lock, it is necessary that we check again, that there is
 no buffer for this lpn.

\begin{algorithm}[htbp]

\caption{FTL write algorithm}

\begin{algorithmic}
{ 
\label{alg:FTLWrite}

\Procedure {FTLWriteSector}{$logical\ sector\ number$}


\State $lpn \leftarrow$ Corresponding logical page number
\State $bufnum \leftarrow$ Search bufLookUpTable for lpn
\If { Search was SUCCESS }
    \State \textsc{WriteLock BUF[bufnum] };
    \State Confirm BUF[bufnum] corresponds to lpn ;
    \State write to BUF[bufnum];
    \State \textsc{WriteUnLock BUF[bufnum] };
\Else
    \State \textsc{ Set Bit in $AllocBufBitmap$ };
    \State Confirm lpn is not in bufLookUpTable;
    \State $bufnum\leftarrow$  Select a Empty/Full/Partially Full Buffer;
      \State \textsc{WriteLock BUF[bufnum] };
     
     \State change bufLookUpTable[bufnum] ;
     \If { BUF[bufnum] is not an Empty Buffer }
     \State  $newtempbuf \leftarrow$ malloc(buffer size)
     \State Swap( $newtempbuf$, BUF[bufnum]) ;
     \State Set flush = 1;
     \EndIf
     \State write to BUF[bufnum];
    \State \textsc{WriteUnLock BUF[bufnum] };
    \State \textsc{UnSet Bit in $AllocBufBitmap$ };
    \EndIf
\If { flush == 1}
\State merge with flash, if BUF[bufnum] was partially FULL;
\State $ppn\leftarrow$ get\_physical\_page()
\State Write $newtempbuf$ to $ppn$ ;
\State Change Maptable ;
\EndIf
\EndProcedure

}
\end{algorithmic}
\end{algorithm}

\subsubsection{FTL Buffer management}

    The buffers are either empty or they could be allocated to some lpn. The buffers which are allocated 
 to some lpn can be either fully dirty or partially dirty. We maintain 2 queues, one that stores buffer 
 indexes for empty buffers and another that stores buffer indexes for full buffers. The buffer indexes
 that are not in empty or full buffer queues are partially full buffers. We attempt to first allocate 
 an empty buffer for an lpn. If that fails then we select a full buffer. If that fails too, then we allocate
 a partially full buffer.  We make use of the RCU based lock free queue implementation from liburcu 
 library \cite{urcu} for the maintenance of buffer index queues.
  
      With the use of lock free queue and by reducing the granularity of locks to the smallest units,
  there is no single central lock contention point in our FTL.

    The evicted full or a partially full  buffer has to be written to flash. For  a partially full buffer,
   we have to do a read-modify-write by merging with the existing content from flash and write to another page.

  We also have  a background thread similar to buffer flush daemon, that  writes the buffers to the flash
 if it is not read or written for more than 60 secs.

\subsubsection{Physical page allocation}

    When we flush a buffer to the flash, a physical page to write to flash is selected. Depending on the
  number of banks, we maintain a `current-writing block' for each bank. Within the current writing block
  the pages are  written sequentially. The get\_physical\_page in  Algorithm 1, selects one of the banks 
  and returns the next page of the `current-writing block' in that bank. The selection of banks is based on
  the activities happening on the bank and is described below.

    Finally, after the evicted buffer is written to flash, the map table is updated.

  The read algorithm is similar. It proceeds by first checking if the logical sector is in FTL cache. 
If not, it reads from the flash the corresponding sector.

\subsection{Garbage collection}

  We have implemented a greedy garbage collection algorithm that acts locally within a bank. 
 We select dirty blocks with less than a minimum threshold of valid pages.
 The valid page threshold depends on the amount of available free blocks in a bank.
 There are three garbage collection levels with different valid page thresholds for  corresponding
 free block thresholds.

  For the first level, the garbage collection algorithm only selects victim blocks with zero valid pages.
 For the subsequent levels the  ``number of valid pages in a block threshold'' is increased. For the first 
 level of garbage collection, there is only erase of victim blocks, that need to be done. For subsequent levels we
 need to do both copying and erase. The greater the level, the greater the copying that need to be done. Our 
 garbage collection algorithm acts locally, in the sense that, the valid pages are copied to another block within
 the same bank. 

Garbage collection can be performed
\begin{enumerate}
  \item before a write request is started: Before servicing a write request, when a new block has to be allocated, we trigger 
the on-demand garbage collection, which creates a fresh supply of free blocks.
 \item  after a write request: In the case of some SSDs, garbage collection is scheduled to happen after the current write
 request is completed. This way, the current write request doesn't suffer the additional garbage collection latencies.
 But if there are any subsequent write requests, those requests are delayed due to the garbage collection overhead.
 \item as a background thread to run in idle time: 
 \end{enumerate}

  We do the garbage collection as separate kernel threads which runs alongside the  FTL I/O  kernel threads.
 Garbage collecting threads try to select a bank for garbage collection which is not being written currently. 
 Similarly when the get\_physical\_page() algorithm selects a bank,  it tries to select a bank that is not garbage
 collected. So the garbage collection is scheduled in such a way that it doesn't interfere much with the I/O.

\subsection{Adaptive Garbage collection}
 Parallelism of the flash card is exploited, when the garbage collection is performed from multiple threads. Each garbage collecting thread 
can work on any one of the bank. With more threads active, more dirty blocks  can be garbage collected. We control the amount of garbage 
 collection performed by controlling the number of simultaneously active garbage collector threads. When the system is relatively idle, 
garbage collection is performed with full force, with all the garbage collector threads being active. When I/O happens, we tone down the 
garbage collection by reducing the number of active garbage collecting threads. During heavy I/O, atmost only one garbage collector
thread is active.

  To implement the adaptive garbage collection, one of the garbage collecting threads take on the \textit{master} role.
The amount of I/O happening in the system is determined by  the number of FTL I/O threads that are asleep and the number of 
threads that are active. So the status of the FTL I/O kthreads are monitored. Depending on the range of
 FTL I/O threads active, we define the range for the number of garbage collection threads that can be active. For example, when all the FTL I/O kthreads
 are active, the number of active garbage collector threads will be one and only the master garbage collector thread will be running. But
 when the I/O in the system drops and fall into a different range, the master garbage collector thread will wake up a few of other
 garbage collector threads. Every garbage collector thread will perform one round of garbage collection on some available banks, before they 
check the status of the FTL I/O kthreads. For the current number of active I/O threads, if the number of GC threads is more than the permissible threshold, 
then a non-master garbage collector thread will put itself to sleep.

  The master garbage collector thread makes sure that a bank with very few free blocks, is compulsorily garbage collected in the near future. 
This is done by flagging the bank as \textit{exclusiveGC} and  this will prevent the get\_physical\_page() algorithm from selecting the bank for writing. 
  This way, we prioritise one of the banks for garbage collection.

\subsection{Checkpointing}
Similar to yaffs2, our FTL writes the in-memory  data structures to the flash when we unload the module and they are read back when it is
 again loaded. We write the main data structures like the maptable, freeblks map, the blkinfo and bankinfo data structures to the flash. 
The blocks that hold these metadata information are differentiated from the blocks that store the normal data by storing a flag in the 
out-of-band area indicating the type of block. When the module is loaded, the flash  blocks have to be scanned for the checkpointed blocks.
This block level scanning can be sped-up by   parallelising the scanning, by reading the flash through several kernel threads during module load.
  
  Actually we don't scan all the blocks, but only a top few and bottom few blocks of every bank. We maintain the checkpoint blocks as a 
linked list on the flash, with one checkpoint block pointing to the next checkpoint block.  We impose a restriction that the first 
checkpoint block is stored in the top few or bottom few blocks of any of the banks in the FTL. So the parallelised scanning only
 reads these few blocks and identifies the first checkpoint block. Since blocks in a bank are written sequentially, we are more 
likely to find a free block in the top or bottom few blocks. In case if none of the banks has a free block on the top or bottom 
few blocks, then we create a free block by moving the contents of a occupied block and write the first checkpoint block there.  

    So during module load time, after finding the first checkpoint block, the rest of the checkpoint blocks, are found by following
 the linked chain.

\section{ Performance evaluation}

In this section, we evaluate the resultant improvements of our parallelism modified algorithms. 
The flash card is mounted in a PCI express slot in HP  workstation machine, with four core intel i5 processors
 running at 3.2GHz. The system has 4GB of memory.  We used the linux kernel version 2.6.38.8 and the operating system
 is Ubuntu 10.04.

\subsection{Number of Queues vs FTL Block device  speed}

  We first give the  measurement over the FTL exposed block device without any filesystem. 
Table \ref{tab:NumQsvsSpeed} shows the measurement of  read and write speed for various number of FTL queues/threads. 
Having a single thread servicing from a single request queue suppresses the hardware parallelism considerably. Increasing the 
number of FTL queues and associated threads we were able to get write speeds upto 360MB/sec.

\ifdefined\THESIS
\begin{table}[h]
      \centering

   \begin{small}
    \begin{tabular}{| l || l || l |}
    \hline
\hline
\textit{\textbf{Number of }} & {\textbf{Read   }} & {\textbf{Write   }} \\  
  {\textbf{  Queues }}  & {\textbf{speed }} & {\textbf{speed }}  \\
   \hline
    \hline
    1        &   53 MB/sec &   45 MB/sec 	   	 \\ \hline
    2	     & 	 99 MB/sec & 	 91 MB/sec		 \\ \hline
    4	     & 	 163 MB/sec & 	 170 MB/sec		 \\ \hline	
    8	     & 	 210 MB/sec & 	 282 MB/sec		 \\ \hline
    16	     & 	 223 MB/sec & 	 364 MB/sec		 \\ \hline
        
    \hline
    \end{tabular}

    \end{small} 
  \caption{Number of FTL Queues vs read-write speed}
\label{tab:NumQsvsSpeed}
\end{table}

\fi

\subsection{Filesystem measurement}

   Table  \ref{tab:wrrdspeed}  gives the filesystem  read and write speed. We mounted XFS filesystem over our FTL.
 The filesystem measurements are taken using tiobench. The results correspond to writing from 16 threads with each 
thread writing 1GB. We see that the existing flash filesystems perform poorly for multithreaded workloads. Yaffs2 for single
 thread was able to get upto 40MB/sec, but with increasing the number of threads, performance deteriorated.  UBIFS was better, but
 still didn't achieve the speeds, the device can support.  We obtain good performance with our FTL used with XFS filesystem.
 We have also shown results, by  using our FTL with the existing mtd\_blkdevs infrastructure to receive block IOs.

\ifdefined\THESIS
 \begin{table}[htbp]

 \begin{minipage}[b]{0.6\textwidth}
      \begin{small}
      \begin{tabular}{|l |l |l |}
\hline    

\textit{\textbf{Filesystem }}  	& {\textbf{Write   }}   & {\textbf{Read   }}  \\  
       {\textbf{  used }} 	& {\textbf{speed }} 	& {\textbf{speed }} \\
   \hline
   \hline
    yaffs2     			 &   5 MB/sec 		&   4 MB/sec 	 \\ \hline
    UBIFS	    		 &   60MB/sec		&   55MB/sec 	\\ \hline
    XFS + FTL using 	   	 &   42MB/sec		&   54MB/sec	\\ 
    mtdblkdevs framework 	 &			&		\\ \hline
    XFS+ FTL using	    	 &   297MB/sec	 	&   245MB/sec	\\ 
    modified multi-queued 	 &			&		\\ 
     framework			 &			&		\\ \hline
    
    \end{tabular}
     
      \end{small} 
\caption{Tiobench results}    
    
    \label{tab:wrrdspeed}
    
  \end{minipage}

\end{table}

\else
 \begin{table*}[t]
 \begin{minipage}[b]{0.6\textwidth}
      \begin{small}
      \begin{tabular}{|l |l |l |}
\hline    

\textit{\textbf{Filesystem }}  	& {\textbf{Write   }}   & {\textbf{Read   }}  \\  
       {\textbf{  used }} 	& {\textbf{speed }} 	& {\textbf{speed }} \\
   \hline
   \hline
    yaffs2     			 &   5 MB/sec 		&   4 MB/sec 	 \\ \hline
    UBIFS	    		 &   60MB/sec		&   55MB/sec 	\\ \hline
    XFS + FTL using 	   	 &   42MB/sec		&   54MB/sec	\\ 
    mtdblkdevs framework 	 &			&		\\ \hline
    XFS+ FTL using	    	 &   297MB/sec	 	&   245MB/sec	\\ 
    modified multi-queued 	 &			&		\\ 
     framework			 &			&		\\ \hline
    
    \end{tabular}
     
      \end{small} 
    
    \caption{Tiobench results}
    \label{tab:wrrdspeed}
    
  \end{minipage}

\end{table*}
\fi

\ifdefined\THESIS

\begin{figure*}[htb]
  \vspace{9pt}

  \centerline{\hbox{ \hspace{0.0in} 
    \epsfxsize=3.0in
    \epsffile{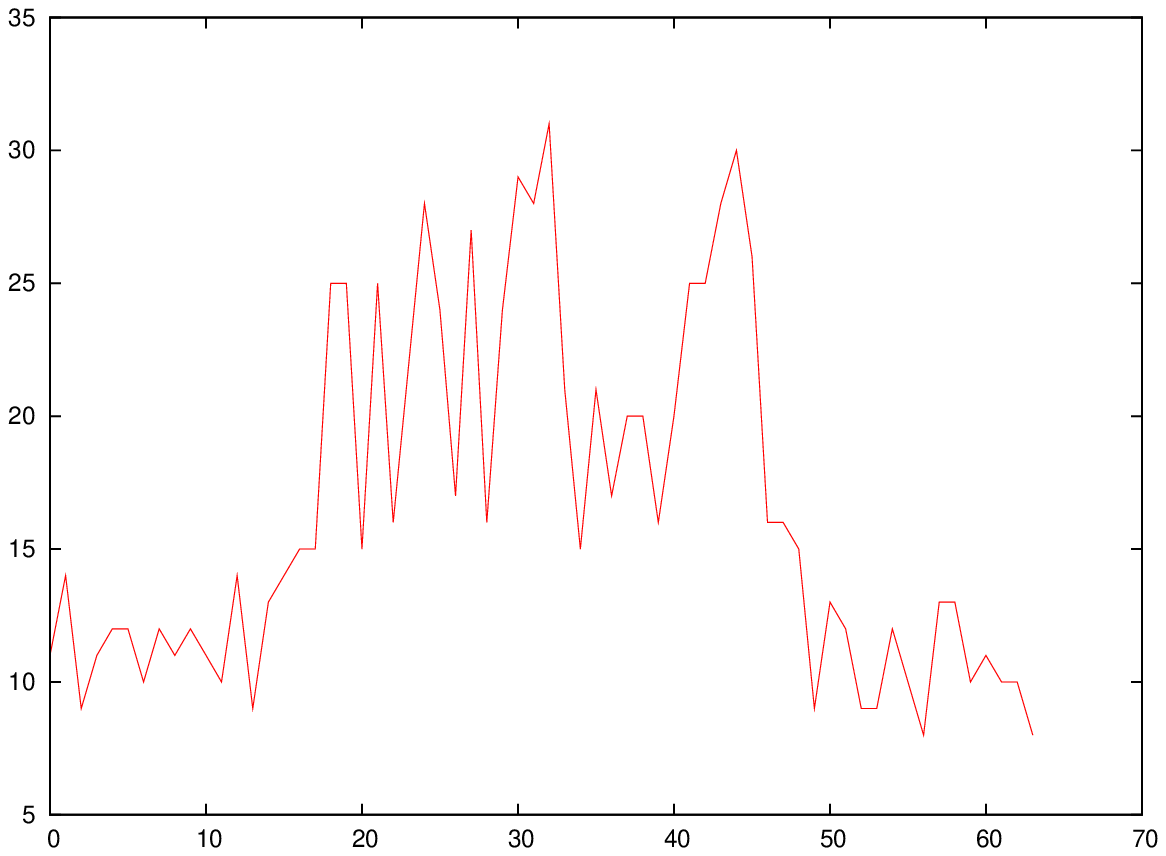}
    \hspace{0.25in}
    \epsfxsize=3.0in
    \epsffile{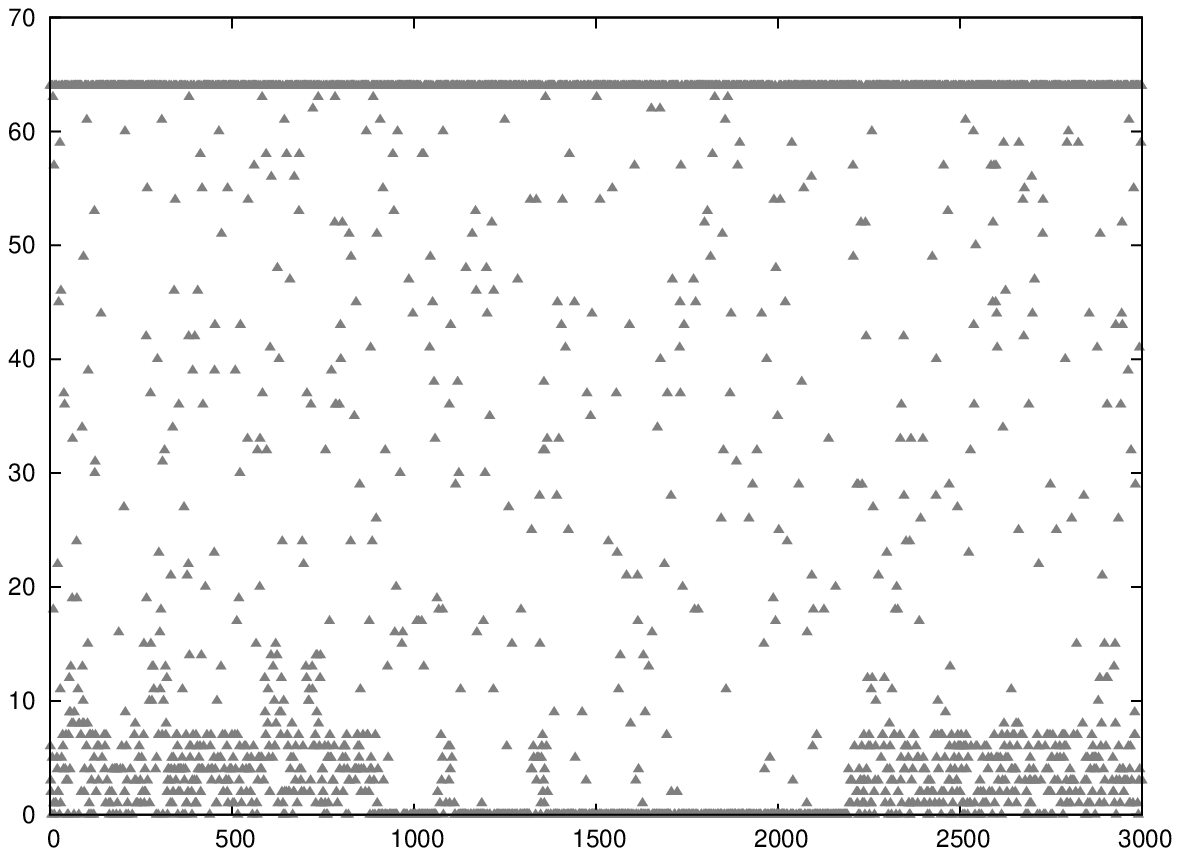}
    }
  }

  \caption{ (a) Distribution of free blocks per bank.
  The x-axis is the bank ids and the y-axis is the number of free blocks in the bank. 
 (b) Distribution of valid pages in occupied blocks.
 The x-axis is the block ids and y-axis is the number of valid pages in the block. }
  \label{fig:freeblkvpagesdist}

\end{figure*}

\else

\begin{figure*}[t]
\begin{minipage}[b]{1.0\columnwidth}
\centering
\includegraphics[width=2in,height=1.5in]{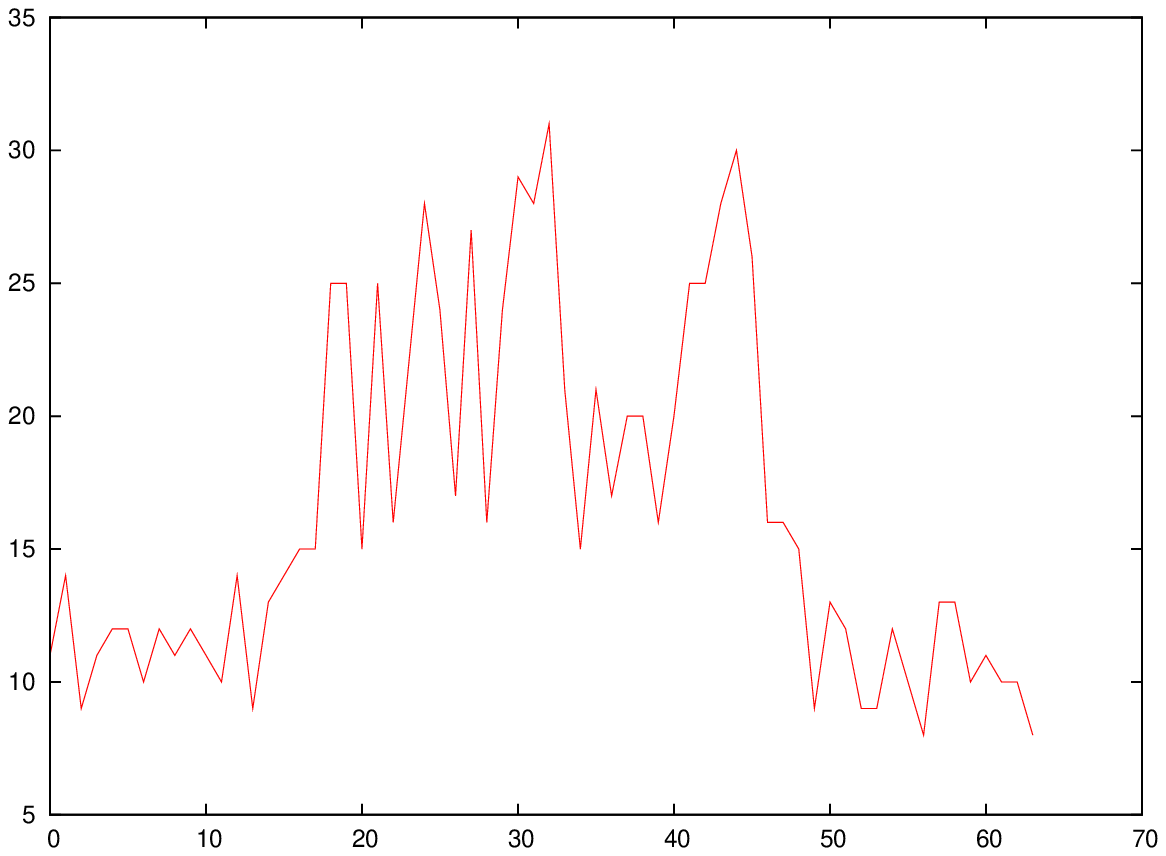}
\caption{ Distribution of free blocks per bank. The x-axis is the bank ids and the
 y-axis is the number of free blocks in the bank}
\label{fig:freeblksdist}
\end{minipage}
\hfill%
\begin{minipage}[b]{1.0\columnwidth}
\centering
\includegraphics[width=2in,height=1.5in]{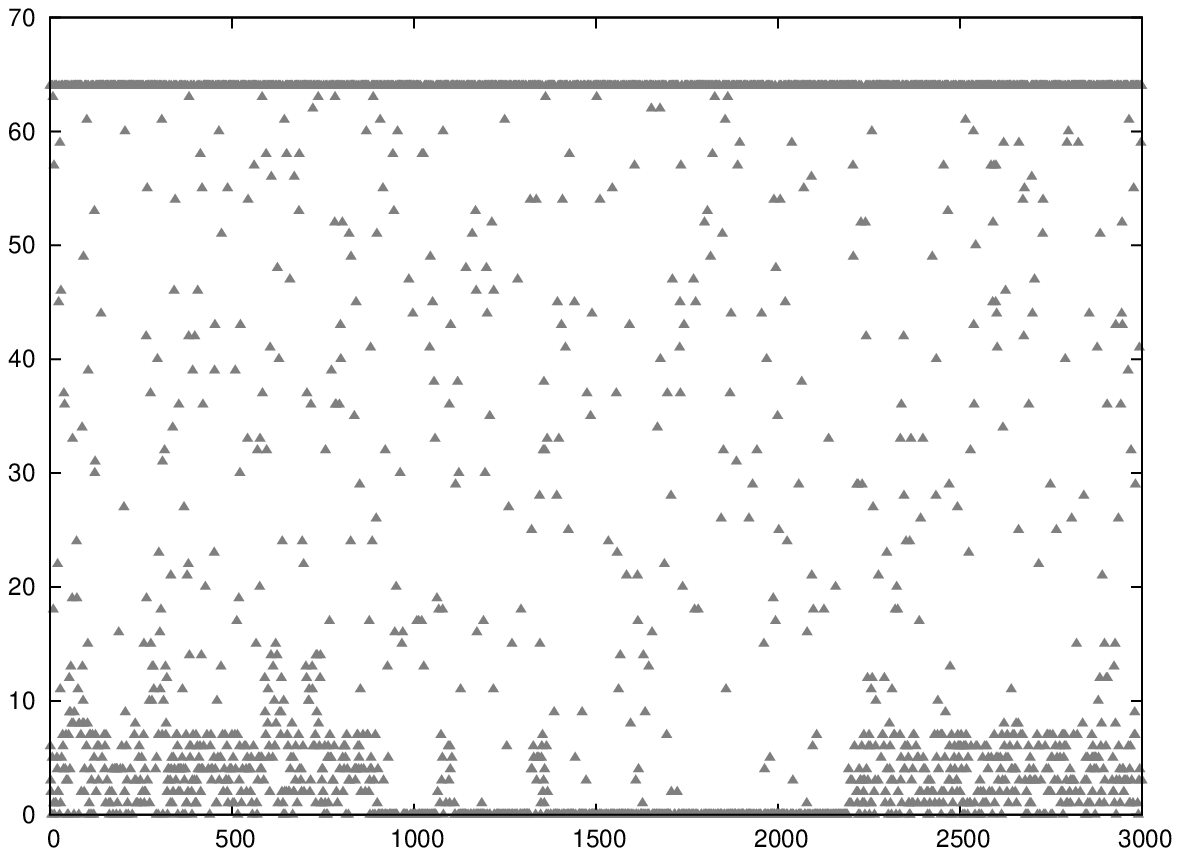}
\caption{Distribution of valid pages in occupied blks}
\label{fig:vpagesdist}

\end{minipage}
\end{figure*}

\fi

\subsection{Module load measurement}

  Table \ref{tab:initscan} gives the initialisation measurements for a 512 GB flash card. We need to do a scanning of a 
few blocks in the beginning of each bank to find the first checkpoint block. As the checkpoint blocks are maintained as a linked
 list on flash, after finding the first checkpoint block we follow the chain. Our FTL load time is only a few seconds. Any mechanism 
that requires a linear block level scan like UBI in 2.6.38.8 kernel,  takes 2 minutes. For Yaffs2, the time taken for module load depends 
upon the amount of data on flash. We observed 16 secs of module load time, when the flash is more than 80\% full.  For failure scenarios, 
for our FTL we need to do a page level scan of the flash. Doing a parallel scan of the pages on the banks still takes 30 minutes. A linear scan will 
 take an inordinately long time, for a few hours.

\ifdefined\THESIS
\begin{table}[htb]
\label{tab:initresultnew}
\centering
   \begin{small}
    \begin{tabular}{| l | l |}
    
\hline

\textit{\textbf{Scanning method}} & {\textbf{Time   }}  \\ 
  {\textbf{   used during module load}}  & {\textbf{taken }} \\
   \hline
    \hline

\hline   
    parallel page 	    &	\textasciitilde 30 min	\\ 
    level scan 	   &   			\\ 
\hline   
    linear block 	    &	\textasciitilde 2 min	\\ 
    level scan of checkpoint	   &   			\\ 
\hline   
    parallel block 	    &	\textasciitilde 28 sec	\\ 
    level scan of checkpoint 	   &   			\\ 
\hline   
    bounded parallel block 	    &	\textasciitilde 2-3 sec	\\ 
    level scan of checkpoint 	   &   			\\

\hline

 \end{tabular}

    \end{small} 
  \caption{Initial scanning time}
 \label{tab:initscan}
\end{table}

\fi

\subsection{Results for Garbage collection}

The garbage collection measurements are measured for three policies:
\begin{itemize}
 \item \textbf{NPGC}: Before a write is done on a bank, if the number of free blocks is less, we garbage collect the
  particular bank, and then subsequently do the write.
 \item \textbf{PLLGC}: Garbage collection is always performed by  parallel co-running garbage collector threads.
  The write algorithm tries, if possible,  to select a bank, which is not being garbage collected. Otherwise the write 
algorithm selects a random bank. 
\item  \textbf{PLLGC+Adaptive}: Similar to PLLGC policy, the garbage collection is always performed by parallel
 co-running garbage collector threads. In addition, the number of garbage collector threads are varied according
 to the amount of I/O. 
\end{itemize}

We use 64 FTL I/O kernel threads for the following measurements.

 First, we compare the the NPGC policy with PLLGC policy. To avoid the long time to start garbage collection, 
we restrict the size of flash as 8GB by reducing the number of blocks per bank to 64. Our garbage collection 
tests were done after modifying the FTL data structures to reflect a long flash usage period.  We artificially
 made the blocks dirty by modifying the FTL's data structures by an IOCTL. The result of the data structure
 modification is shown in Figure \ref{fig:freeblkvpagesdist}(a) and Figure \ref{fig:freeblkvpagesdist}(b).
 The Figure \ref{fig:freeblkvpagesdist}(a) shows the distribution of free blocks per bank, which is roughly normally
 distributed. The Figure \ref{fig:freeblkvpagesdist}(b) shows the distribution of valid pages in a block.  After the
 modifications of the data structures, the garbage collection measurements were taken.  The maximum number of GC thread
 is set to 1 for this measurement. 

\ifdefined\THESIS
\FloatBarrier
\fi

\begin{table*}[htb]
\ra{1.3}

\caption{Comparison of garbage collection done before a write and  parallel garbage collection }
\centering

\begin{tabular}{| l| l | l | }
\toprule
	    & NPGC policy  & PLLGC policy  \\ \midrule

Scatterplot of write latencies & \includegraphics[width=2.3in,height=2in]{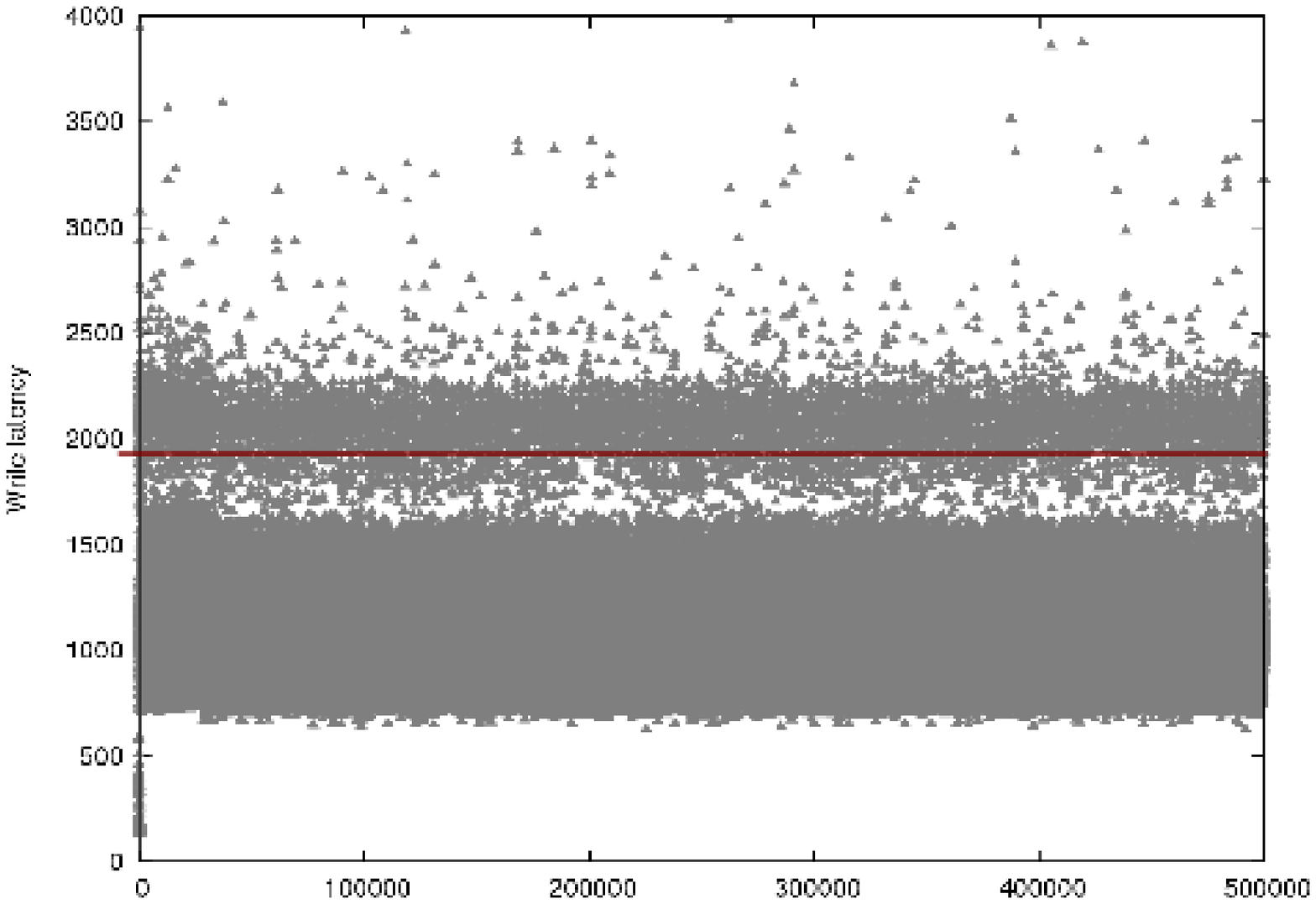}&\includegraphics[width=2.3in,height=2in]{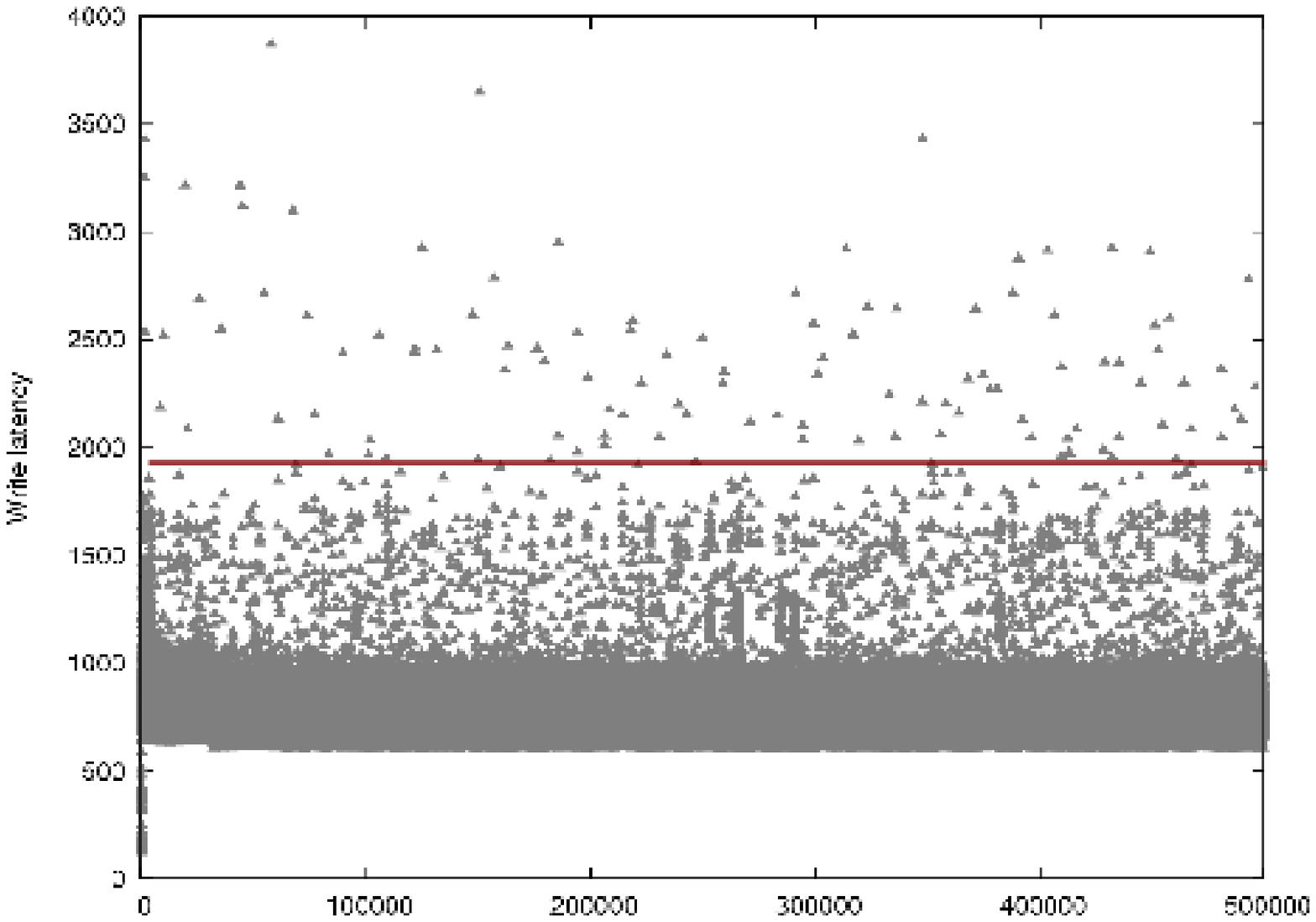} \\ \hline

Number of blocks & 8312 blocks & 8392 blocks \\ 
garbage collected &     &      \\ \hline
Elapsed time & 546 sec & 400 secs \\ \hline

\bottomrule
\end{tabular}
\label{tab:compNPGcPLLGc}
\end{table*}

    For this measurement, we do over-write of 1GB of data upto 16 times from a single threaded application.
 We show the latencies of write requests from the single application thread. The application program does a 
fsync after every 4K to ensure that the data reaches the FTL device.  The Table \ref{tab:compNPGcPLLGc} compares
 the NPGC and PLLGC policy showing the  scatter-plot  of write latencies, for every written 32KB, in microseconds.
 The x-axis in the scatterplot is the request ids and the y-axis is the write latency for that request.

 The scatter plot shows the number of requests that cross the 2ms boundary is much more in NPGC policy because of the GC overhead. 
 When we do the garbage collection in parallel, the GC overhead is considerably removed.  It is also important to notice  that
 we were able to garbage collect  roughly  the same number of blocks without incurring the additional GC overhead, by scheduling
 the garbage collection activity on a parallel bank. The elapsed time is also shorter in the PLLGC version.
   The completion of the test took 546 seconds when garbage collection is done in the write-path  and only 400 seconds for PLLGC policy.

    Next, we provide the measurements comparing  PLLGC with PLLGC+Adaptive policy for the scenario of  multithreaded applications.
 We set the  maximum number of garbage collector threads to be 8 and we did the evaluation by also reducing the number of banks 
 on the flash to 8. We used a simulated workload with 128 application pthreads having a fixed think time. Each application pthread
 sleeps for 20ms after every 32K write and 10sec after every 2MB write and each pthread writes 4MB of data 16times. So in total, we 
 write 8GB of data and we measured the average write latency as observed from each pthread. The plot in the Figure \ref{fig:normavglat} shows
 the normalised average latency as seen from the 128 application pthreads. The adaptive version performed better than the non-adaptive parallel
 garbage collector. Also the total elapsed time for the test was 217 secs for PLLGC+Adaptive policy while PLLGC policy  took 251 secs to completion.

\ifdefined\THESIS
\begin{figure}[hb]
  \vspace{9pt}

  {\hbox{ \hspace{0.0in} 
    \epsfxsize=3.4in
    \epsffile{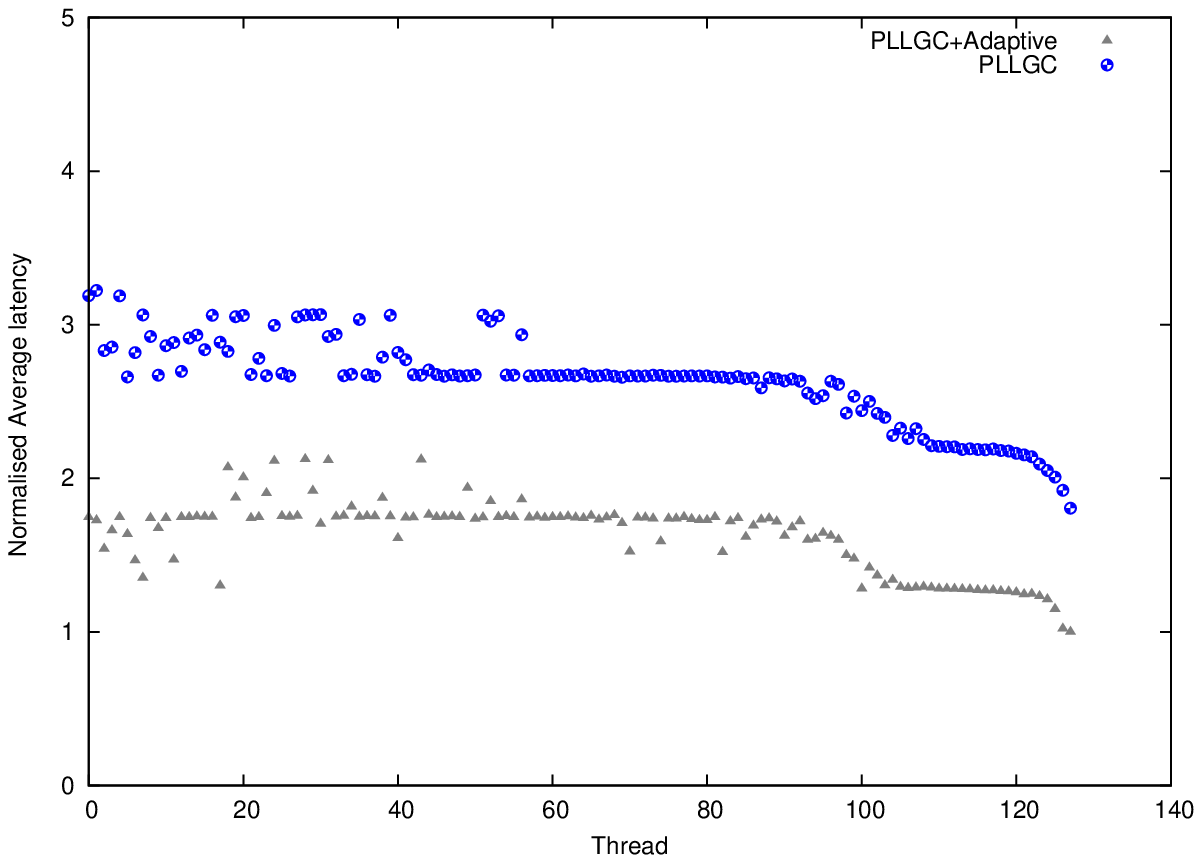}
    \hspace{0.25in}
    
    }
  }

  \caption{ Normalised Average write latency for the 128 threads. 
The x-axis is the thread ids. For every thread we calculate the average latency.
 The plot is made after normalising with respect to the minimum average latency.}

\label{fig:normavglat}

\end{figure}
\else
\begin{figure}[h]
\begin{center}
 \includegraphics[scale = 0.5]{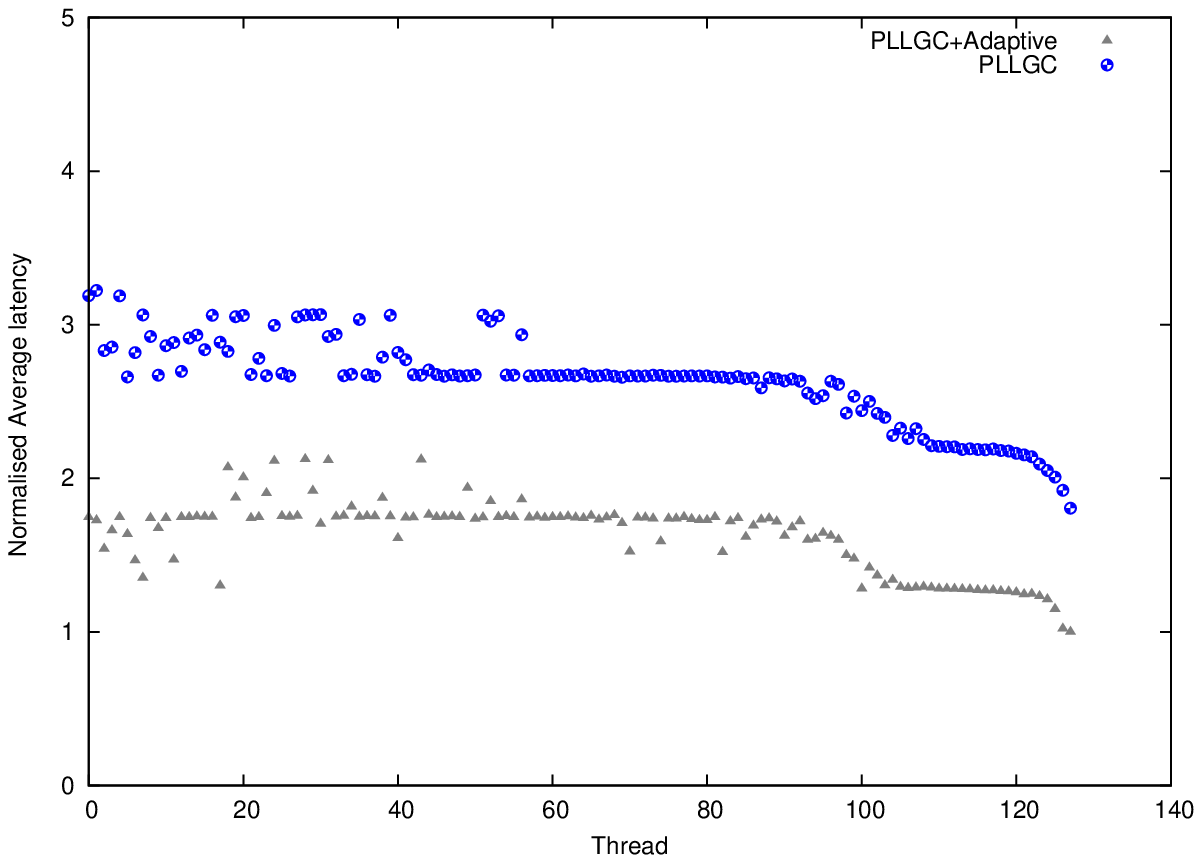}
 
\caption{ Normalised Average write latency for the 128 threads. The x-axis is the thread ids. 
For every thread we calculate the average latency.
 The plot is made after normalising with respect to the minimum average latency.}
\label{fig:normavglat}
\end{center}
\end{figure}
\fi
 
\section{ Related Work}

 The use of multiple queues per device in the block layer to achieve greater speed has been discussed in the linux community \cite{lwnmq}. 
Bypassing the  block layer request queue management by registering a new make\_request is also done for SSDs by \cite{driverPCIeSSD}. 
 They have also proposed to bypass the SCSI/ATA layer of the I/O stack.

  Scheduling the garbage collection activity according to internal SSD parallelism is discussed in \cite{PreemptiveGC}. 
Here the garbage collection is split into phases of page read,write and block erase. They define pre-emption points between 
these phases and service write requests in between. Using the multi-plane command, the garbage collection read/write are allowed 
to happen  concurrently with normal reads and writes. 

  In our work, we try to exploit the  parallelism (interfaces, banks, chips) offered by the raw flash card, that we used. 
  In our work, we have also considered an adaptive method to vary the amount of garbage collection that is happening at a particular instant.

Recent works in parallelism with respect to SSDs are \cite{exploit||ism10}, \cite{SSDllism11}, \cite{B+treellism} , \cite{IOScheduler}.

Scheduling write requests by the flash controller in an SSD in order to exploit the parallelism is considered in \cite{exploit||ism10}. 
They have proposed a few  heuristic algorithms in order to load-balance the IO requests across the device queues.

In \cite{SSDllism11}, they have given performance results on SSDs  for different access workload in the context of parallelism.
 Their measurements show that because of parallelism read speeds  can be slower than write speeds in SSDs.
In \cite{B+treellism}, parallelism is exploited for B+tree usage.
In \cite{IOScheduler}, I/O scheduling strategies for SSDs are considered with exploiting parallelism as one of the goals.

\section{ Discussion }

   With each written page on the flash, we store some metadata in the out-of-band area(spare bytes).
   We store the type of block (blocks that store checkpoint data or normal data), the corresponding logical page number,
   and a sequence number. The  sequence number is similar to the one used in yaffs2 and is used for failure
   scenarios, when the FTL module is  unloaded incorrectly. After an incorrect unload, the FTL has to do a
   page-level scan of the flash. Failure handling is important in battery operated embedded devices. But there
   are several  usecases, other than embedded systems, for the kind of Raw Flash cards, that we used.  For failure
   handling, we also should implement the block layer barrier commands \cite{IObarrier}. When the filesystem gives
   a barrier request, we have to empty the queues in our FTL and we  also have to flush all of our FTL buffers to 
   the flash.
 
    The FTL cache size is a module parameter and it is limited only by the amount of RAM in the system. So if we
    expect very heavy random write workload, we can load the FTL module with larger cache. It is also possible that
    if we can detect random IO behaviour, the FTL can adaptively increase the FTL cache size. We can also increase or
    decrease the per buffer size. This kind of tunable control is not possible with using the SSDs directly, but it is 
    possible with using the raw flash card and a software FTL.

   We have been liberal in the use of memory in our FTL. But it is possible to reduce the memory consumption considerably.
   The bits necessary for bit locking in our write algorithm can be stolen from the maptable entries. Though we have used a 
   fully page based FTL in our implementation, it is possible to extend the purely page based FTL with DFTL\cite{DFTL} 
   algorithm and reduce memory consumption of the mapping table.

    There are many directions for future work. In addition to doing the  garbage collection in parallel, it is also possible to 
  do other activities like prefetching, static wear-levelling or de-duplication in parallel without affecting the write latencies.

    We used  a FIFO buffer replacement in our FTL, which is not ideal when there is temporal locality in the workload. One future work
   is to consider other locality aware cache replacement policies. It would be interesting to consider how the non-blocking buffer
   management schemes like \cite{Nbgclock} compare against conventional LRU schemes.

    In our design of the FTL, we have avoided using the block layer's request merging, coalescing and IO scheduling optimisations.
    One direction of future work is to look in to the scheduling policies, that can be used in our multi-queued design.  It has been shown
    that by allocating blocks based on hotness of data, we can improve garbage collection efficiency and hence the long term endurance of the 
    flash. We have to explore further if the selection of banks in which the blocks are allocated, should also be based on hotness.

\section{ Conclusion}

In conclusion we have shown how a FTL can be designed for a
multi-banked flash card.  The modifications leverage the parallelism
available in these devices to achieve higher  performance.  We 
show that garbage collection overhead can be considerably removed from
 the write path because of the available parallelism. In addition to using 
parallelism for greater speed, we also show that initial scanning time that 
is required in flash can also be reduced due to parallelism.  We have also proposed
 an adaptive garbage collection policy, to vary the amount of garbage collection, according
 the IO load, by controlling the number of active garbage collector threads.

\section{ Acknowledgements}
Thanks to authors of open source  flash filesystems(Yaffs2, UBIFS and logfs), for mailing list discussions. Special thanks to Joern Engel.

{\bibliographystyle{abbrv}
  \bibliography{lftlbib}
}

\end{document}